# Information centric networking over SDN and OpenFlow: Architectural aspects and experiments on the OFELIA testbed

S. Salsano[(1)], N. Blefari-Melazzi[(1)], A. Detti[(1)], G. Morabito[(2)], L. Veltri[(3)]

(1) University of Rome "Tor Vergata" / CNIT, Rome (Italy)
(2) University of Catania / CNIT, Catania (Italy)
(3) University of Parma / CNIT, Parma (Italy)

**Abstract** - *Information Centric Networking* (ICN) is a new networking paradigm in which the network provides users with content instead of communication channels between hosts. *Software Defined Networking* (SDN) is an approach that promises to enable the continuous evolution of networking architectures. In this paper we propose and discuss solutions to support ICN by using SDN concepts. We focus on an ICN framework called CONET, which grounds its roots in the CCN/NDN architecture and can interwork with its implementation (CCNx). Although some details of our solution have been specifically designed for the CONET architecture, its general ideas and concepts are applicable to a class of recent ICN proposals, which follow the basic mode of operation of CCN/NDN. We approach the problem in two complementary ways. First we discuss a general and long term solution based on SDN concepts without taking into account specific limitations of SDN standards and equipment. Then we focus on an experiment to support ICN functionality over a large scale SDN testbed based on OpenFlow, developed in the context of the OFELIA European research project. The current OFELIA testbed is based on OpenFlow 1.0 equipment from a variety of vendors, therefore we had to design the experiment taking into account the features that are currently available on off-the-shelf OpenFlow equipment.

# 1. Introduction

The shift from "host-centric" networking to "Information Centric" or "Content-centric" networking has been proposed in several papers (e.g. [1][2]) and is now the focus of an increasing number of research projects ([3][4][5][6][7][8]). In short, with Information Centric Networking (ICN), the network provides users with content, instead of providing communication channels between hosts. The network becomes aware of the content that it is providing, rather than just transferring it amongst end-points. As a result, it can introduce features such as in-network caching or content-based service differentiation. The biggest advantages of ICN can become reality when network nodes natively support ICN mechanisms. Deploying ICN capable equipment into existing networks is a critical issue as it may require the replacement or update of existing running equipment. In this context, we believe that Software Defined Networking (SDN) is an important opportunity as it promises to enable the continuous evolution of networking architectures. SDN is an architecture characterized by a logically centralized control plane and a well-defined separation between user and control planes [9][10][11]. SDN devices implement actions and rules decided by a possibly remote controller. SDN abstracts the network like the operating system of a computer abstracts applications from its hardware. SDN could lead to a software era of networking, providing a built-in capacity for evolution and countless innovations.

This is the first important motivation to investigate how ICN functionality can be supported using SDN. An SDN enabled network could facilitate the introduction of ICN functionality, without requiring re-deployment of new ICN capable hardware.

A second motivation for the integration of these two paradigms can be found in our specific approach to ICN. The framework we have defined, called CONET (COntent NETwork) [12] [13], is based on the interaction of the forwarding nodes with nodes belonging to a routing layer (called NRS: Name Routing System). We believe that such architecture can help in overcoming some critical issues of ICNs, such as scalability. In fact, one of the major problems of ICN is related to the fact that the number of content items (and therefore, names) is expected to be orders of magnitude larger than the number of hosts. Therefore, performing name-based routing in each node in the path between source and destination could be extremely costly in terms of processing, which results in increased delay and reduced throughput. To reduce the impact of such problem CONET confines routing to a subset of nodes, that is NRSs. The forwarding nodes are instructed by the NRS nodes and this approach of separation between the forwarding plane and a "control plane" naturally maps into the SDN architecture.

The most popular implementation of the SDN paradigm is OpenFlow [9][10][11]. In OpenFlow, network nodes are called OpenFlow Switches and are responsible for forwarding, whilst routing decisions and control functions are delegated to a centralized element called "Controller". OpenFlow Switches communicate with the Controller through the OpenFlow protocol. The OpenFlow protocol is already implemented in a number of commercial products, and has been the focus of several research and prototyping projects. Started as a research/academic project, OpenFlow evolution and standardization efforts are now covered by the industry forum named the *Open Networking Foundation* (ONF) [11].

In this paper, we investigate the support of ICN using SDN concepts from two different perspectives: i) a medium-long term perspective, with no reference to the limitations of current OpenFlow specifications and available equipment; ii) a short term perspective, in which we restrict ourselves to OpenFlow v1.0 specification and equipment and describe a prototype implementation running on the OpenFlow v1.0 testbed provided by the OFELIA project [14].

In Section 2, we provide the main concepts of the CONET ICN solution, summarizing our previous work on this topic. The proposed SDN-based solution to support ICN, in which the Controller nodes can process events coming from forwarding nodes and can instruct the forwarding nodes both in a reactive and in a proactive way is described in Section 3.

In Section 4 we propose the CONET ICN packet format which is suitable for the SDN/OpenFlow approach and analyze the issues related to the flow matching capabilities in OpenFlow switches. With reference to the long term perspective for ICN support over SDN, the set of supported operations and the needed capability of forwarding nodes are discussed in Section 5, while Section 6 discusses the "northbound" API that should be exposed by the Controller nodes. In the definition of the proposed solutions, we start from the latest OpenFlow specification (v1.3) but we assume that we can arbitrarily enhance it and that we can have equipment capabilities matching our ICN needs.

In Section 7, we consider the "short term" solution to support ICN using the OpenFlow v1.0 specification and equipment. This solution has been fully implemented and tested in the OpenFlow v1.0 testbed provided by the OFELIA project. We report in the paper the results and the experience we gained from the implementation.

Finally we point out that although some details of our solution have been specifically designed for the CONET architecture (e.g. to take into account the CONET packet format) its general ideas and concepts are applicable to a class of recent ICN proposals, which follow the basic way of operation of CCN/NDN. Thus, the applicability of our proposal for supporting ICN by using SDN concepts is not limited to our own CONET architecture but is far more general.

## 2. The CONET ICN Solution

When considering the integration of SDN and ICN, we took a pragmatic approach and started from a specific ICN solution called CONET. CONET is based on the concepts introduced in Content Centric Networking/Named Data Networking (CCN/NDN) architectures [2],[3]. It extends the approach proposed by CCN/NDN in several aspects, including integration with IP, routing scalability, transport mechanisms, inter-domain routing. From an implementation viewpoint, we started from the CCNx [8] implementation and kept compatibility with CCNx based applications, as we offer the same API. In this way we can reuse in our experiment the existing CCNx applications and support them using CONET protocols. In this section we describe our previous work on CONET, as needed to understand the combined ICN/SDN that will be described in the following sections.

Figure 1 shows an overview of the CONET architecture, which has been first proposed in [12]. The "terminals" are called *ICN Clients* and *ICN Servers*. ICN Clients request content using CONET protocols, using the name of the content to be retrieved. ICN Servers are the originators/providers of the content. A terminal can act as both ICN Client and ICN Server if needed. Using the same terminology as CCN [2], the content requests are called "interests" whereas the related packets are called "interest packets". The interest packets are forwarded over the ICN network, taking into account the requested content-name. When a node that contains the content receives the interest packet, it replies with a "data" packet that is sent back towards the requesting node. Data packets follow back the path towards the requester and intermediate nodes can store the content, performing transparent "in-network" caching. Therefore further interests for the same content can be served by intermediate nodes rather than by ICN Servers.

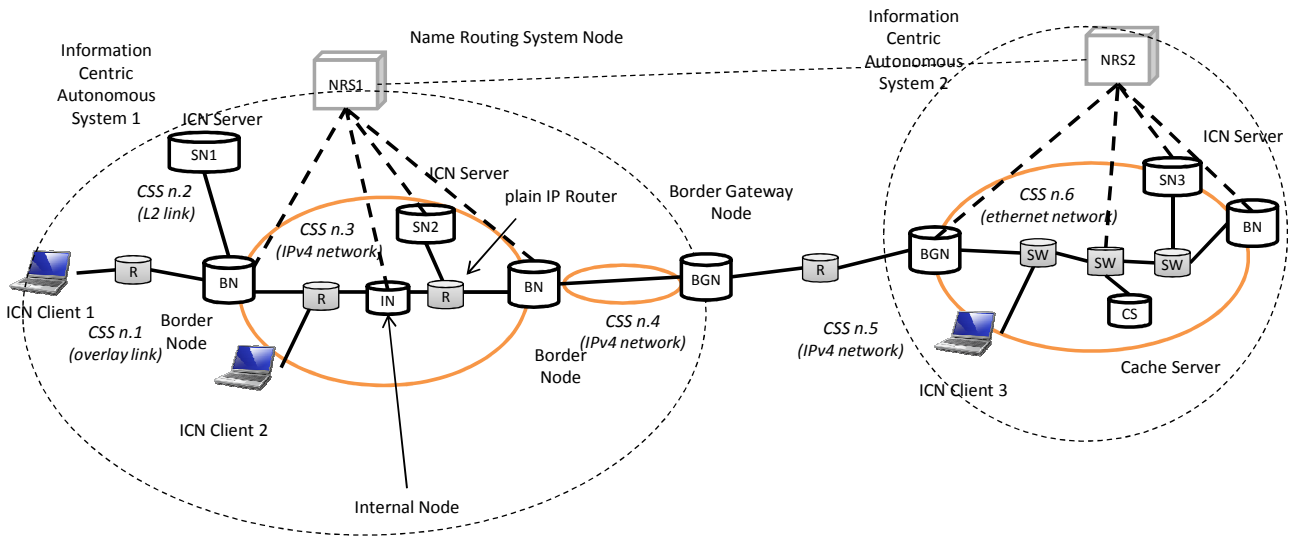

**Figure 1: CONET Architecture**

A set of logical components interwork in an ICN solution, such as *naming*, *forward-by-name* of interests, *content routing* (how to disseminate information about location of contents), *data forwarding*, *in-network caching*, *segmentation and transport*, *security*. We will only give a high level introduction of the different aspects, making reference to the published documentation.

As for the *naming* we support the naming proposed by the CCN/NDN architecture. The *content name* is a string, such as: foo.com/football. It can be human readable, as in this example. We can see it as the composition of a *principal* (foo.com) and a *label* (/football) [1]. Such a naming scheme allows us to aggregate routes and to perform longest prefix matching. We also support *self-certifying names* [1], by using a public key as the string of the principal. Therefore CONET supports both human-readable and self-certifying names.

*Forward-by-name* of interests consists in a name-based lookup table and on a prefix matching algorithm. When interest packets arrive, the name-lookup operation has to be performed at line-speed, using a table that stores the associations between name prefixes and next hop. This table is called FIB (Forwarding Information Base), like in IP routers. Moreover, another table, called RIB (Routing Information Base) is needed to exchange routing information with other nodes and it does not need to be accessed at line speed. The RIB and FIB could have tens of billions of entries in order to include all the possible content names, making it infeasible to implement them in router hardware. Therefore in CONET we use the FIB as a cache of currently needed routes and employ a centralized routing logic that may serve a set of forwarding nodes, managing the full routing table (RIB). We have called this approach "Lookup-and-Cache", a typical sequence of forwarding operations is shown in the upper half of Figure 2. The forwarding node N1 receives an interest for "foo.com/text1.txt/chunk1, segment 1". Since the FIB lacks the related route, the node temporarily queues the interest and looks up the route in a remote node that manages the RIB, called *NRS (Name Routing System) Node*. Node N1 gets the next-hop info, stores it in the FIB and forwards the interest packet. The detailed design and scalability implications of the Lookup-and-Cache approach have been analyzed in [13]. We addressed two scalability concerns: the size of the RIBs and the rate of the lookup requests on the interfaces between the forwarding nodes and the NRS. Considering a workload comparable to current http content requests over backbone links, we have verified the system scalability.

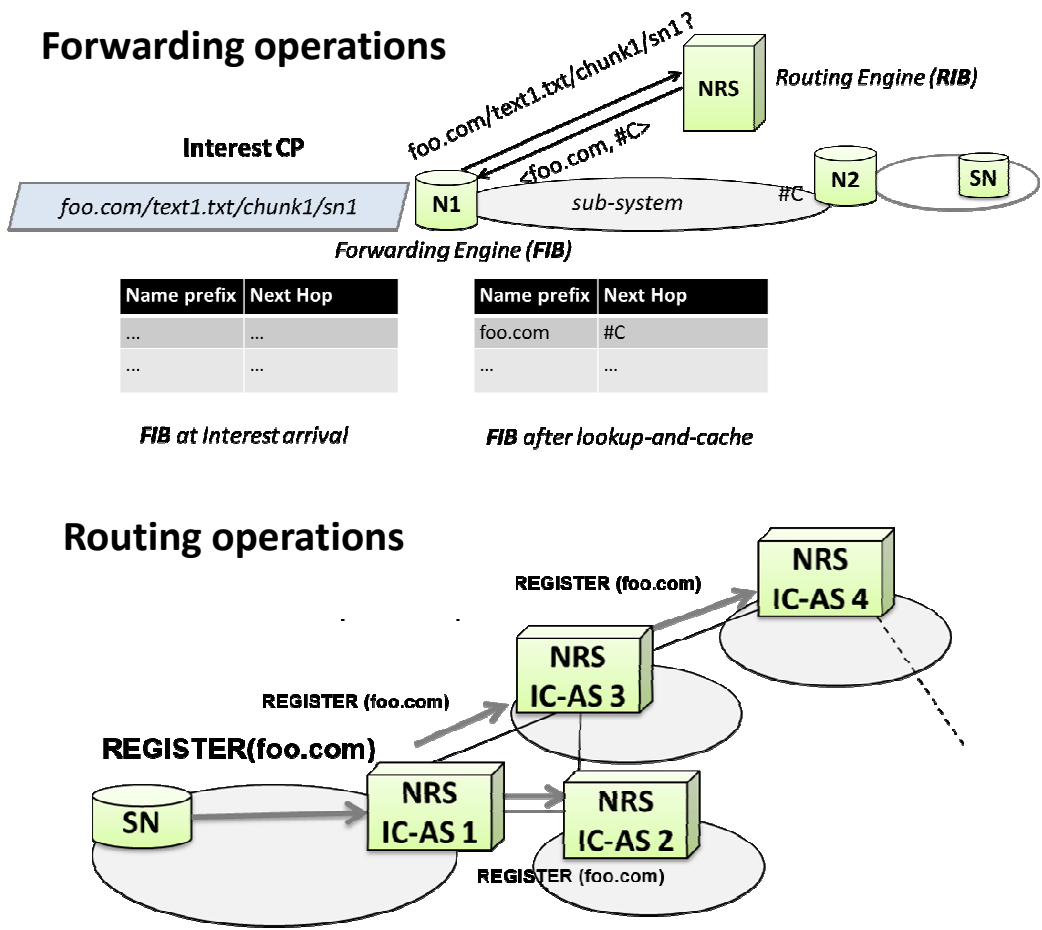

**Figure 2- Lookup and cache: forwarding and routing operations**

The NRS nodes are responsible for the *content-routing* mechanisms, both at the intra-domain and inter-domain level. The NRS functionality can be seen as logically centralized within one autonomous system. All name-based routes are contained in a RIB, which logically serves all nodes of a sub-system. Thus, the expensive Routing Engine can be logically centralized rather than replicated on all the forwarding nodes. Scalability and reliability issues will obviously drive the deployment of NRSs in a distributed and redundant way, but these aspects are not yet addressed in our prototypes. The proposed Name Routing System has some similarity with the Domain Name System, as it provides the resolution of a content name into a next-hop, while the DNS provides the resolution of a domain name into an IP address. Apart from this aspect, the functionality of the NRS is not based on the DNS.

The proposed CONET inter-domain routing approach is conceptually similar to the current BGP inter-domain routing, in which the different ASs advertise the reachability information of a network prefix, along with the AS path used to reach that prefix. This information exchange is performed among the NRS nodes of the different ASs, as shown in the lower part of Figure 2. Our current prototype implementation disseminates name-prefixes using "REGISTER" and "UNREGISTER" messages, with a mechanism similar to the one that has been proposed in the DONA architecture [1]. The scalability of this approach to name-based inter-domain routing is a research topic on which we are still working. As discussed in [15], such scalability concerns for inter-domain operations are common to all ICN approaches in the literature.

As for the *segmentation and transport* issues, the content to be transported can be very variable in size, from few bytes to hundreds of Gigabytes. Therefore, it needs to be segmented into smaller data units, typically called *chunks*, in order to be handled by ICN Nodes. A chunk is the basic data unit to which caching and security is applied. In CONET, we handle the segmentation of content with two levels: at the first level the content is segmented into chunks, at the second level chunks are segmented into smaller data units (called *Carrier-Packets*). The transfer of Carrier Packets is regulated by our proposed "Information Centric Transport Protocol" (ICTP) [16] [17]. ICTP performs reliability and congestion control, much like TCP does for the transfer of a byte stream in current TCP/IP networks. The proposed segmentation mechanisms allow using chunks bigger than the layer 2 Maximum Transmission Units (MTU) without incurring IP level fragmentation, which happens for example in the CCNx [8] implementation of CCN. In [16] the performance merits of the proposed solution are discussed.

Existing implementations of ICN solutions, such as CCNx [8], mostly rely on an "overlay" approach to run on top of existing IP based networks. This overlay approach works by tunneling ICN information units within TCP or UDP flows running over IP. We proposed an "integration" approach for ICN design that evolves from current IP technology and tries to extend it towards ICN support. In particular, we have proposed a way to carry ICN related information in IP headers by defining new IP options [18] for IPv4 and IPv6. This information is processed only by a subset of ICN capable nodes, which are capable of performing forward-by-name and ICN data forwarding operations (BN and BGN in Figure 1), while legacy routers simply forward the packets looking at regular IP header information.

## 3. Support of CONET ICN in SDN

Following the SDN approach, we consider an OpenFlow-based ICN architecture in which the intelligence of the ICN is de-coupled from the forwarding (of interest and data packets) and caching functions. As shown in Figure 3, this results in an architecture that is composed by two different planes: i) a data plane with the ICN Servers (i.e. the content producers), the ICN Clients (i.e. the content requesters/consumers) and the ICN Nodes; ii) a control plane that includes the Name Routing System (composed by NRS Nodes), a security infrastructure (e.g. a PKI - Public Key Infrastructure) and "Orchestrator" nodes. The two planes communicate through an *extended OpenFlow* interface, used by the NRS nodes (acting as OpenFlow Controllers) to control one or more ICN Nodes. Note that the introduction of NRS Nodes in the CONET architecture discussed in Section 2 pre-dates the integrated ICN/SDN solution here discussed, but it is fully aligned with the SDN approach of using "controllers" to drive the forwarding behavior of switches/routers and to enforce an explicit separation between a data forwarding plane and a control plane. In this architecture, the name-based routing intelligence needed to implement ICN functionality runs in the NRS, implemented as a set of OpenFlow controllers. In the control plane the controllers/NRS Nodes offer also an API towards the "Orchestrator" node that orchestrates the overall behavior of a domain. This is referred to as the "Northbound" interface in the SDN jargon.

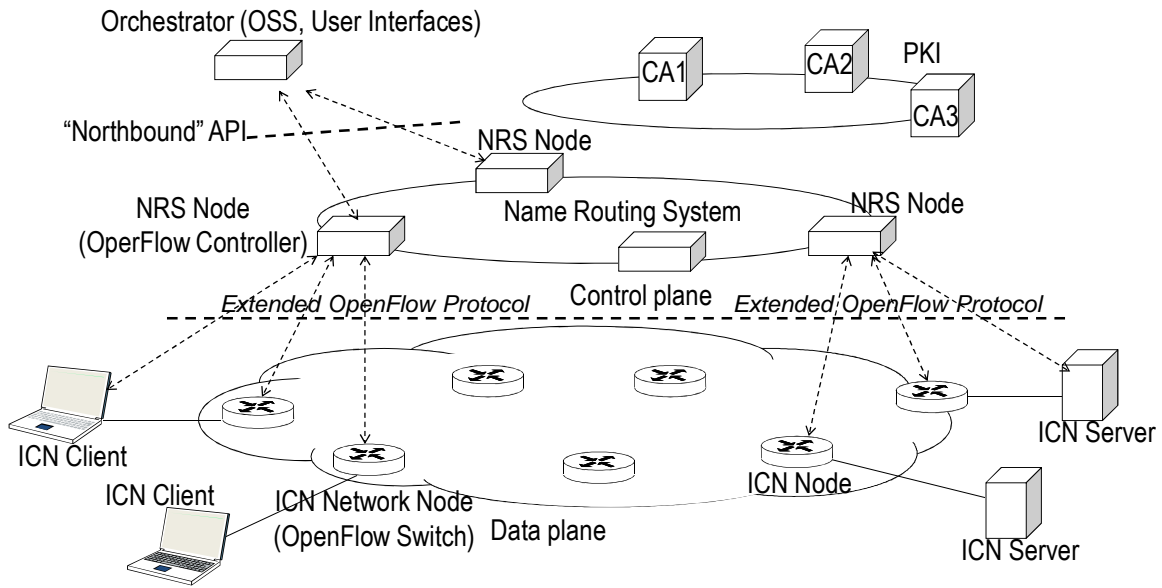

**Figure 3- Information Centric Network based on extended OpenFlow**

Figure 4 shows the set of reference interfaces for the design of our solution. The interface iN lies between the NRS Node and an ICN Node. Figure 4 shows two NRS Nodes, and represents a scenario with two domains (two IC-ASs). The two NRS nodes are interconnected by the iC interface, which is not an OpenFlow interface, its definition is out of the scope of this document. The interface iS is placed between the ICN Server and the NRS node. This interface is used by the ICN Servers to publish into the NRS the information related to the hosted contents. We have designed the interface iS as an extended OpenFlow interface. We assume that an NRS Node/OF controller offers the iM interface towards a "Management" node playing the role of orchestrator and offering User Interfaces towards network managers. In our work we implemented the iM interface extending the REST API provided by the Floodlight OpenFlow controller.

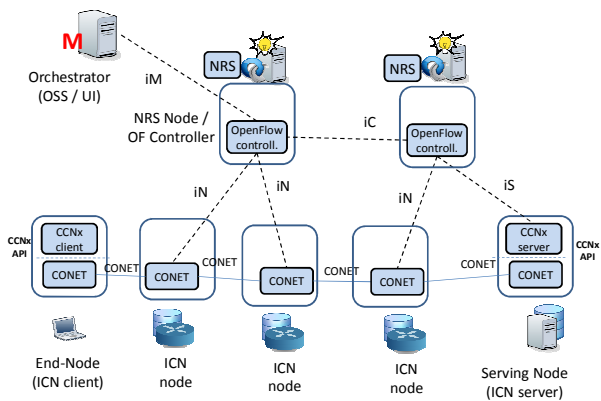 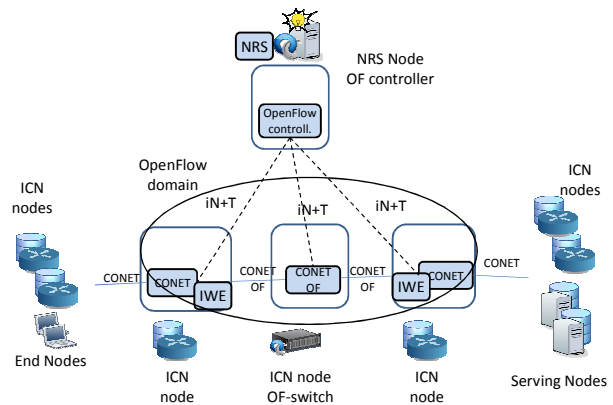

**Figure 4- Reference Interfaces**             **Figure 5- Reference Interfaces for intra-domain specific SDN functionalities**

In Figure 4 we assume that a common CONET protocol is used throughout the ICN network (and across different domains) and that the ICN Nodes can exploit the information carried within the CONET protocol to perform the required operations. On the other hand, within an OpenFlow based domain, we may want to add further information that can be processed by OpenFlow capable equipment within that specific domain. This information needs to be removed outside of the domain. We refer to this process as "Tagging", we added "InterWorking Elements" (IWE) within ICN "ingress/egress" Nodes and we denote with iN+T an extension of the iN interface that also provides this functionality (see Figure 5).

## 4. CONET packet format and matching in OpenFlow switches

In our early design of CONET ICN [12], we proposed to use an IP option to carry ICN related information (e.g. the content name) in IP packets. Including information in the IP header was meant as an elegant and "clean" way to enhance the networking layer with ICN functionality. In the context of the exploitation of SDN functionality we observe that the same result can be achieved by forwarding nodes looking at information carried by higher protocol layers (e.g. transport layer). Therefore we propose here an alternative solution that does not even require the definition and use of the CONET IP Option, but carries the ICN information in the transport protocol layer.

Let us first consider the ICN related information used by an ICN Node to process a packet (further details of all header fields and their encoding are reported in [18]). Each interest packet is targeted to a specific content chunk and must include both the content name, called ICN-ID, and the Chunk Sequence Number (CSN). The ICN-ID can be of fixed or variable size, according to the selected naming approach, and may support different naming schemas. The CSN may be included as part of the content name or added explicitly as variable-length optional field. A one byte length field called "Diffserv and Type" has been also defined in order to differentiate the quality of service operations that can apply to the content. It can also identify the class of content (e.g. "live video" or "file"). These header fields are present in both interest and data CONET packets. As described in previous sections, Data chunks are further segmented into a set of Carrier Packets that are individually encapsulated within CONET packets. This ICN related information is needed by ICN Nodes on a packet by packet basis, when forwarding (or caching) decisions are taken.. Using the ICTP segmentation and transport described in Section 2 we carry ICN related information in the headers of all packets, so that ICN Nodes can properly handle them[1]. We also take advantage of these headers carried in all IP packet for OpenFlow based operation, as shown hereafter.

The different ways of transporting name-based information within IP that we have analyzed are depicted in Figure 6. The IP protocol field (IP PROTO), which identifies the IP payload (e.g. TCP or UDP) is set to a value that identifies the ICN (CONET) payload type[2]. Our early proposal was to exploit the IP options field of standard IPv4 and IPv6 headers, defining a new CONET IP option (#1 in Figure 6), and carrying the rest of the Carrier Packet as payload. The novel proposal (#2 in Figure 6) is to include the ICN information as transport adjacency to the IP header, similar to IPSec for IPv4 (or as IP header extensions in IPv6). The processing of information that follows the basic IP header is in line with current operations of "middle-boxes" like NAT and Firewalls and with the SDN/OpenFlow architecture, in which flow table entries can match on protocol headers from layer 2 up to layer 4 in a cross-layer fashion. The advantage of the first solution is mostly conceptual: it allows an ICN Node to take routing decision by looking only at information contained in the layer 3 (IP) header. Major disadvantages are: legacy IP nodes could have some problems with unrecognized IP options (experiencing higher processing times or even dropping such packets); the implementation in end nodes is more complex as it requires changes in the IP layer. In [12] we investigated (with practical experiments on PlanetLab) how our unrecognized IP option is handled by current routers in the Internet. In the large majority of tests it was possible to add unrecognized options and achieve end-to-end CONET connectivity among arbitrary PlanetLab nodes, while in few cases some routers in the path have dropped the packets.

---

[1] Using CCNx overlay based implementation, the chunks are carried in UDP packets, which gets fragmented at the IP level. The IP fragments that are generated lose the needed ICN information. A CCNx node can only operate reconstructing the IP fragments.

[2] The IP PROTO code for CONET transport protocol should be assigned by IANA after an IETF standardization process, an experimental value can be used (codes 252 and 253 are reserved for experimentation and testing).

The advantages of the second proposal are complementary: the implementation in terminals is simpler and legacy IP nodes are not affected in any way by the ICN information carried in the IP payload. Even assuming that the few problematic routers could be reconfigured to solve the issues with the first packet format, we believe that the format #2 is to be preferred for these advantages. In fact, the disadvantage of using transport layer information to process the packet does not apply to a node with an SDN/OpenFlow architecture. Considering more "traditional" router architectures, the need of processing transport layer information is a rather typical requirement considering that most routers also implement some middlebox [19] functionality. An ICN Node can be seen as a middlebox capable of performing ICN routing (and caching) functionality. In this context, we note that the identification of ICN packets by means of a specific code in the IP PROTO field represents an efficient solution.

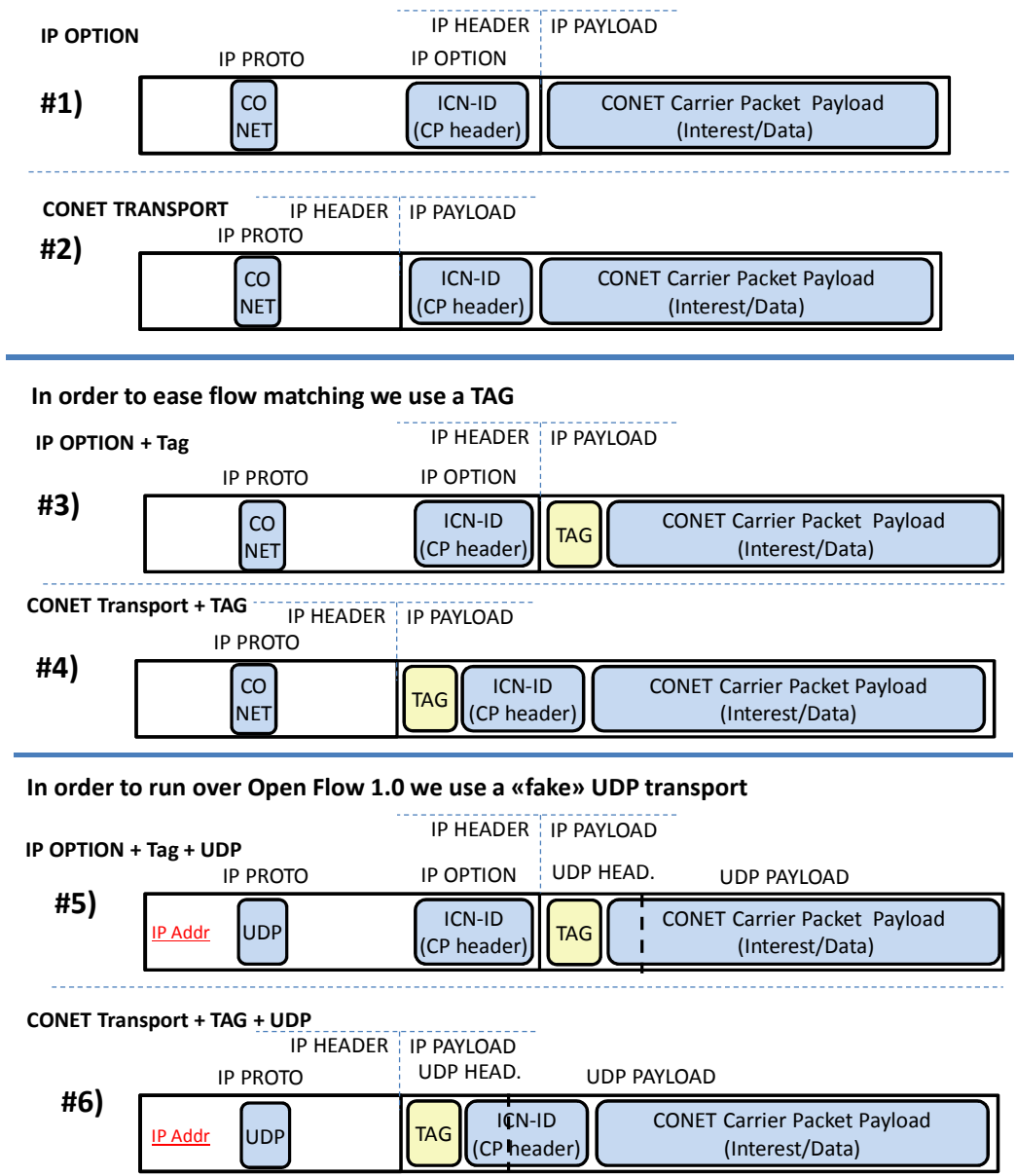

**Figure 6- Different choices for CONET packet format**

### 4.1 Packet matching and its critical aspects

Let us now consider how an OpenFlow based ICN Node can process ICN packets. OpenFlow allows the insertion in the switch flow table of standard table entries, formed by a matching rule based on standard layer2, 3, and 4 fields, and the corresponding forwarding action. However, as new capabilities and packets formats are defined, extensions to the matching capability of OpenFlow nodes are needed. The OpenFlow 1.2 specification has introduced "flexible matching" for new fields and protocol types. These can be successfully parsed and matched with proper flow table entries. Albeit, this does not enable OpenFlow 1.2 (or the current 1.3 version) to parse a generic unknown protocol field. Rather, it just grants the possibility to extend the OpenFlow protocol, in a modular way, such that it becomes aware of new header fields.

Take for example the extension required to match the packet format #1 described in Figure 6. In this case, an OpenFlow switch needs to match ICN information inserted as IPv4 or IPv6 options header fields. This is still not possible in current OpenFlow 1.3 specifications. In order to match the new IP option code, a new "Flow Match Field" code can be added to the OpenFlow protocol (and the OpenFlow switches need to be extended to support it). If the IP option type matches the new defined CONET IP option, further matching has to be performed against the ICN ID field within the CONET IP option in order to provide different actions for different contents. However, this goes beyond the current OpenFlow match capabilities, as the ICN-ID field is a variable length field. Its length can be in the order of tens of bytes, for example we can (rather arbitrarily) fix a maximum length of 128 bytes. We could think of an OpenFlow protocol enhancement in future releases, but the capability to match a variable length field of such dimension could still remain problematic to implement at hardware level.

In case of the packet format #2 described in Figure 6, the flow table rule for CONET packet can start with the matching against the "Protocol" field of the IP header, specifying the chosen CONET IP protocol number. This is clearly simpler than the matching of IP option. However, when passing to match the specific ICN-ID value, we face the same issue previously described (variable length field).

### 4.2 The tagging approach

We experienced that matching an arbitrary name of variable length, as required by ICN is fairly complex from the point of view of an OpenFlow switch, and probably it is not reasonable to expect that this capability will be implemented in hardware. One could think of using fixed length expressions for naming content, solving the problem at its root. Unfortunately this may not be adequate for most naming schemas and can limit the flexibility of the ICN design. For this reason we further propose to dynamically map content names to fixed length tags in "ingress" nodes at the border of an ICN/OpenFlow domain. The nodes that perform this mapping will be called Edge Nodes (EN). The tags can be used within a domain and can be removed by "egress" Edge Nodes. In principle, the tags could be carried in data units of existing protocols, re-using some existing fields. For example VLAN tag or MPLS label could be (ab)used for this purpose. This approach has the disadvantage that a coexistence mechanism should be considered if the used technology is also used with its regular semantic. Therefore, to avoid collisions with other uses of pre-existent tagging technologies, we have also defined a way to explicitly carry the tag within the CONET protocol. Considering application of this approach to both the proposed packet formats #1 and #2, this respectively leads to the packet formats #3 and #4 in Figure 6, where a tag is added at the beginning of the CONET header. In our design we considered a 8 bytes tag. Differently from the original packet formats #1 and #2, it is now possible to operate with fixed length match over the tag. The needed extension to OpenFlow protocol and to the switch capabilities are straightforward in this case as they are just a typical example of the "flexible matching" mechanism introduced with OpenFlow v 1.2.

Packet formats #5 and #6 in Figure 6 will be discussed in Section 7, which deals with our implementation over the OFELIA testbed based on OpenFlow v1.0 equipment.

## 5. ICN related OpenFlow operations

In order to handle the concept of "content" and to support ICN-related operations some extensions of the current OpenFlow protocol are required in terms of new supported methods and exchanged data. We consider here a "long term" vision and assume that we can arbitrarily extend the OpenFlow protocol according to the ICN needs. An initial description of the issues related to this approach was first presented in [20]. We have now started an implementation of the proposed long term solutions, using software based switches (Open vSwitch), but we do not yet have mature results, therefore this section is mainly conceptual and does not make reference to the concrete implementation that will be discussed in Section 7.

New ICN-related operations can be classified, and hereafter summarized, as: "content-related", "routing-related", "tagging-related" and "security-related".

### 5.1 Routing-related operations

An ICN adopts an addressing scheme based on names, which do not include references to their location; as a consequence, interest messages have to be routed toward the closest copy of the content, based on the content-name. This introduces the need of routing-related operations that are:

*Forwarding-by-name* – When receiving a new interest message, the ICN Node is expected to forward it to a next-hop based on a name-based forwarding table present within the node (the FIB described in Section 2). The insertion of new entries in such forwarding table is the responsibility of the NRS Nodes and should be controlled through the OpenFlow iN interface. Particularly, if an interest is received for a content name that is not in the local forwarding table, the node requests name-to-next-hop lookup to the controller that implements the NRS functionality, as described in section 2. Particular care should be taken when the forwarding table is full and a previous entry has to be deleted and replaced by the new one. Such operation (and all relevant decisions such as which entry to delete) may be assisted by the controller, or can be autonomously taken by the node on a basis of fixed and simple rules.

*Forwarding table management* – Forwarding table can be also populated or modified asynchronously and proactively by the NRS Nodes according to some upper level strategy (for example by distributing the forwarding-by-name information for the most popular contents, or for contents that require some form of prioritization).

*Forwarding table exportation* – In an instance where the NRS Node does not keep a copy of the forwarding table of each controlled node, such information should be provided by the ICN Node on demand.

### 5.2 Content-related operations

*Content publications* – An ICN Server publishes through the NRS all content names that it is capable of serving.

*Caching decisions* – An ICN Node may provide caching functionality for achieving a native, in-network caching function. This may drastically improve multicast content distribution (when the same content has to be distributed from one source to multiple destinations), and more in general, allows for a more efficient content delivery in both fixed and mobile environments [21] (when the same content is successively requested by other destinations). Due to the large amount of packets (and content) that a node is requested to forward, it is expected that a node decides to cache only a portion of the overall forwarded contents. The decision of which content is to be cached, how long, and which is not, can be made locally inside the node or, more likely, by relying on the NRS Node.

*Caching notification* – It could be expected that the NRS Node is notified when a particular chunk of content has been locally cached by a controlled node.

*Proactive caching* – The NRS Node may proactively push content to an ICN Node anticipating the "automatic" caching procedures that occurs on data transfer. These "proactive push" operations could prove very useful in the case of live content distribution within the ICN (e.g. audio/video real-time streaming).

*Interest handling* – It could be useful to keep a trace of requested contents (e.g., to optimize the processing of further interest requests). This can be done locally in the ICN Node or delegated to the NRS Node, performed on a per-request basis (every time a new interest message arrives) or ii) on a batch basis (sending to the controller periodic summary information reports about the requested content).

## 5.3 Tagging-related operations

*Name-to-tag mapping–* In case tag-based routing is performed within a single domain, an ICN Node should be able to request a tag for a given content to the NRS Node, and the NRS Node could asynchronously "inject" a name-to-tag mapping into an ICN Node.

## 5.4 Security-related operations

*Security enforcement* – Content (or content chunks, like in CONET) are cryptographically protected in order to assure content (and content generator) authenticity and data integrity [22] through a digital signature with the private key of the content (or of the content generator). Every ICN Node should verify such a signature before forwarding the content toward the interested ICN Clients.

*Key management and distribution* – For self-certifying names, the association between names and public keys is embedded in the name. For other naming schemas, this association should be maintained by the NRS, according to a proper key management mechanism, and provided by the NRS Node to its controlled ICN Nodes.

*Key revocation* – In parallel to a suitable key management and distribution mechanism, a key revocation mechanism should also be implemented, to allow the revocation of compromised or withdrawn keys.

## 5.5 Design of OpenFlow protocol extensions

In order to support such new operations, proper OpenFlow protocol extension has been studied. In OpenFlow, exchanged messages are encoded in binary format. Each message is assigned a code, listed in the ofp_type enumeration (section A.1 of [23]). The code reserved for "Experimenter" messages is exploited in our ongoing implementation of the extensions. With this message type, we introduced a "generic" extended message which can convey operations and parameters encoded with a JSON syntax. In this way we can encode new extension messages using a JSON syntax within standard binary OpenFlow messages. This leads to an interface that can be more readily extended. In case further implementations want to favor message length rather than readability, an equivalent binary format can be defined for each extension message.

The specific messages we proposed to support the CONET/ICN operations are described in detail in [24]. For space constraint we only list the message categories here: Cached content notifications, Content publication, Name to next hop messages, Connection setup, Name-to-tag mapping.

## 6. SDN Northbound API for ICN

In the context of SDN, the Northbound API is primarily meant to offer application developers a more powerful API to exploit SDN capabilities. The definition and standardization of this API is currently a hot topic in the SDN research community. Contrary to the OpenFlow API, which has been the main object of standardization so far, the definition of the Northbound API is still at a very early stage. In this context we believe that it is worth considering also ICN functionality that could be offered over this interface.

As we used the Floodlight controller [25] in our implementation, we started our consideration from the Floodlight Northbound interface, based on a HTTP REST interface. The Floodlight interface offers the capability to gather several information from a controller, for example regarding the discovered network topology and the status of the flow tables configured in the switches.

We added the capability to query for the list of cached content items stored in each cache node, so that the Orchestrator can learn where the different content objects are available. As a practical tool, we also added the capability to activate and deactivate the ICN caching in a specific node.

As further extensions we think it would be very useful to define commands to gather statistics on the requests for the different content objects by the users and to populate the cache nodes with content in advance. The definition and implementation of these capabilities is in our next plans.

## 7. Experimentation in the OFELIA testbed

OFELIA [14] is an European research project aimed at creating a large distributed testbed for SDN based on OpenFlow. During the lifetime of the project the testbed is open to academia and industry for running experiments with OpenFlow. The OFELIA testbed [26] offers network virtualization, and lets different experimenters run their experiments in parallel. OFELIA consists of autonomous OpenFlow enabled islands (currently 9) that offer their services and facilities to experimenters through a "Control Framework" (CF). The OFELIA CF is derived from the Expedient tool developed at Stanford University in the eGENI project [27]. Using the CF an experimenter can create its own "slice" i.e. virtual topology over the set of physical OpenFlow switches, can instantiate its own OpenFlow controller(s) and a set of "user" Virtual Machines that may play the role of traffic sources/sinks.

We used the OFELIA testbed for experimenting with the solution described in this paper, as was first discussed in [28]. We consider a scenario in which an OpenFlow based network (i.e. the OFELIA testbed) is a part of a larger ICN network (Figure 7). Within this OpenFlow based network, the delivery of content exploits the specific features of this domain. Whilst, outside the OpenFlow network, non-OpenFlow ICN processing will be used. Such a general scenario includes the case in which the OpenFlow domain corresponds to the entire network. The compatibility between the OF domain and the external world is assured by the IWE in the Edge Node (EN) which translates regular ICN operations to OpenFlow based ICN operations and vice-versa. Our main assumption is the use of OpenFlow 1.0 switching equipment (i.e. off-the-shelf switches that cannot be enhanced).

We considered a scenario in which content can be stored and returned by ICN Cache Servers. In order to keep compatibility with existing hardware, we use Cache Servers external to the OF switches. In the future, the caching functionality should be integrated into the switches. Note that although the communication between the controller (i.e. the NRS Node) and switching equipment is based on OpenFlow 1.0, the communication between the controller and the external "Cache Server" does not need to be OpenFlow v1.0 compliant.

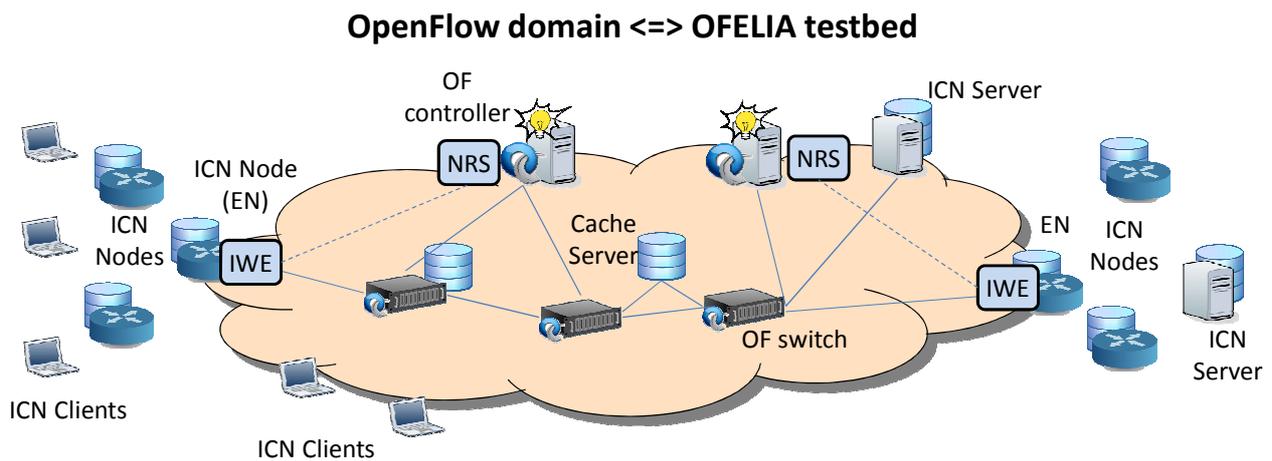

**Figure 7- Details of the OpenFlow 1.0 solution to support ICN**

Let us now describe the proposed operations. An Interest packet is received by an EN (sitting at the border of the OF domain) and the content has not yet been cached by any node in the OF domain. If the content requested by the Interest has originated from an ICN Server internal to the OF domain, the Interest packet will be forwarded to such an ICN Server, otherwise the interest packet is forwarded towards the outgoing EN on the path to the target ICN Server. If the content has been already cached within an "in-network" Cache Server, the Interest packet has to be forwarded towards it.

When the data packets flow from the ICN Server towards the receiving host, a traversed OF switch will duplicate the data packets and send a copy to the Cache Server.

In our assumptions, the OpenFlow domain can be a single IP subnet, operating at layer 3. Alternatively, it may be a set of IP subnets, interconnected by OF capable routers. In both cases, the incoming EN resolves the ICN name into the IP address of the ICN next hop (i.e. the ICN Server or the outgoing EN), using ICN routing information and forward the packet through the OF domain. Likewise, the ICN Server or the outgoing EN will send back the content Data packets to the incoming Edge Node through the same domain. With OpenFlow 1.0 compliant equipment it is not possible to read and process the content name (ICN-ID) within the ICN packet, thus we chose to implement the forwarding function within the OF domain based on a tagging operation (tag-based routing). The mapping from content name to tag is performed by ICN ENs with the help of a NRS node, which is logically centralized for a single OF domain, ensuring that the mapping from name to tag is unequivocal across the OF domain. When removing the tag, the outgoing EN will also cache the content-name- to-tag association, so that when the content data packet returns, it can add the tag and forward the tagged packet in the OF domain.

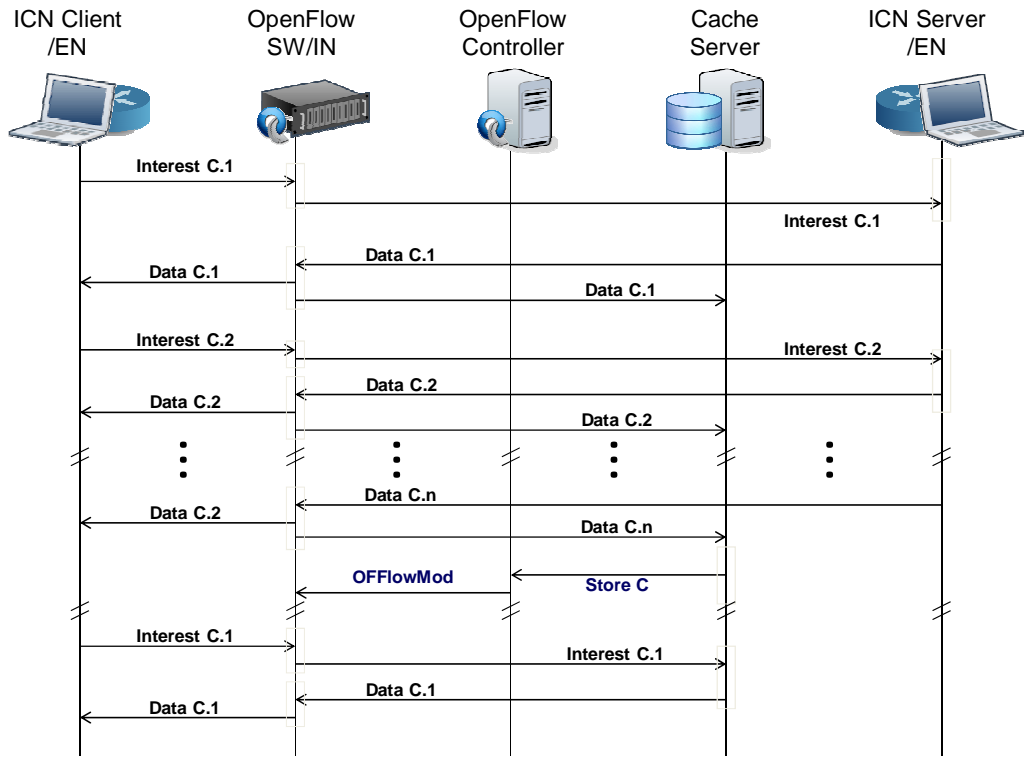

**Figure 8- Sequence of operations**

Figure 8 shows the sequence of operations in the OF domain. As the data packets return, they are duplicated towards the Cache Server. When the Cache Server has cached a full chunk, it notifies the controller, which adds the flow information to the switch. Further interest packets for the same chunk of content may also be directed to the Cache Server. It is responsibility of the NRS to decide whether the interest packet should be forwarded towards the ICN Server or the Cache Server.

In order to work with OpenFlow v1.0 equipment we defined a solution with different packet formats with respect to the ones described in Section 4. This solution belongs to the category of abusing an existing field of a protocol to carry the tag. In particular we reuse the UDP source and destination port to carry the tag, according to the packet formats shown as #5 and #6 in Figure 6. The tag is carried as the first four bytes of the IP payload, corresponding to the source and destination port header fields in the case of UDP and TCP transport protocols. We selected protocol type 17 (UDP) to identify CONET packets, as OpenFlow v1.0 only allows the inspection of transport protocol header for UDP and TCP packets. Therefore the tag can be inspected by OpenFlow v1.0 switches as the source and destination port of a fictitious UDP header. In order to distinguish real UDP packets from CONET packets, we reserve an additional IP address on ICN terminals for CONET use. In this way, the combination of IP destination addresses and UDP ports (containing the tag) unambiguously identifies an ICN content. An example of this addressing scheme is shown in Figure 9, which reports the layout of the deployed testbed. The OF switches identify the packets with IP address 192.168.1.8 as CONET interests and the packets with IP address 192.168.1.23 as CONET data. Note that the use of these special addresses is only due to the limitations of the OpenFlow 1.0 API and existing switches.

Note that the Cache Server can be associated one-to-one with an OpenFlow switch (and in this case a direct link between the two nodes can be used), or they can be "remotely" located considering a one-­to-many relation between a Cache Server and a set of OpenFlow switches.

Figure 9 shows the first simple configuration implemented over the OFELIA testbed, involving the ICN Client, the ICN Server, the Cache Server, the OF Controller realized using Floodlight [25] and two OpenFlow switches. The ICN Client and ICN Server run the default CCNx applications (ccnrepo, ccnputfile, ccncatchunks2) and interact with our CONET enhanced version of the ccnd daemon. The client and server also implement the IWE in order to send CONET packets, including the tag that is used to map the content name.

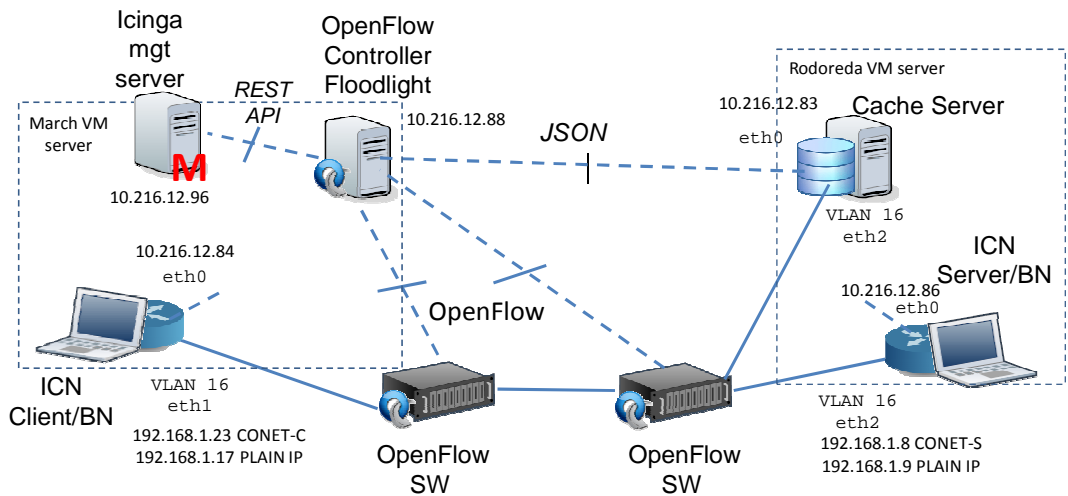

**Figure 9- ICN testbed in OFELIA**

Some results from the testbed are reported in Figure 10, which shows the amount of traffic traversing the network interfaces of the ICN Client, ICN Server and Cache Server and the number of cached items in the Cache Server. For the network interfaces of the three types of nodes the incoming and outgoing traffic are plotted in Figure 10: the (green) filled histograms represent the incoming traffic and the (blue) empty histograms the outgoing traffic. In the experiment, more than 200 different content chunks are continuously requested by a request generator running on the ICN Client. Looking at the ICN Client load, the outgoing load corresponds to the interest requests, the incoming load corresponds to the content data, therefore it is considerably higher. The experiment is divided in three phases as highlighted by the vertical dashed red line. In the first phase, the ICN Server provides the content, as the ICN caching is disabled. The load in the ICN Server is symmetric with respect to the load on the ICN Client: interest packets are seen as incoming and content data are seen as outgoing. As indicated in the figure, during this phase the controller is instructing the switches to work as plain Ethernet "MAC learning switches". In the second phase, the caching is enabled, by changing the control logic operating in the controller. This is done operating on the web interface of the management server shown in Figure 9. The management server uses the controller REST API to change the control logic. The controller installs the rules in the switches so that the ICN data packets are also copied toward the Cache Server, which can start caching the chunks. In turn, the second phase is divided into a transient phase and a steady-state phase. In the transient phase, the number of cached items increases until all chunks are cached. In this phase, both incoming and outgoing load on the ICN Server decreases as more interest requests for cached contents are redirected towards the Cache Server. The incoming load on Cache Server increases in the transient phase, because copies of the data chunks are sent toward the Cache Server. In the steady state phase all chunks are cached and the load on the ICN Server decreases to zero, as all the client requests are served by the Cache Server. In this phase, the incoming load on the Cache Server only includes the interest packets. The Cache Server has fully replaced the ICN Server, this is due to the fact that in our experiment a predefined set of 200 different files is used as content to be requested. In the third phase, the ICN mechanism is disabled by sending a command to the controller. The controller clears all the content related entries in the switches and instruct the switches to operate again as plain MAC learning switches. All the requests are now forwarded towards the ICN Server and its load is again as high as in the first phase.

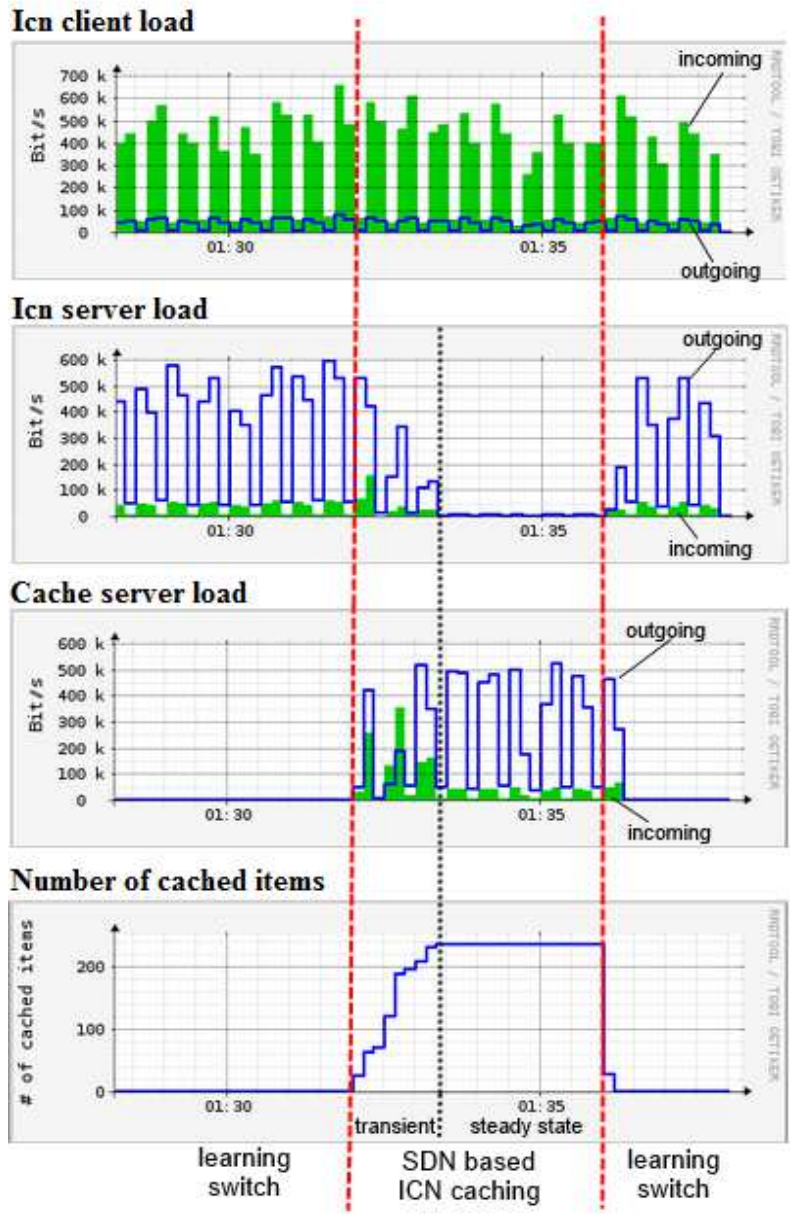

**Figure 10- CONET experiment in OFELIA OpenFlow 1.0 testbed**

## 8. Related Work

The work on the relation between SDN and ICN is relatively recent. Besides our previous work ([20][28]), we are aware of the research activities presented in [29]. Authors of that paper exploit the SDN concepts for improving the design of their ICN solution [4][30].

In [31] the authors propose ICN as one use case of their "SDIA - Software Defined Internet Architecture". The described operations of domain edge nodes share many points in common with our approach, being managed by an edge-controller that instructs the edge nodes to insert the appropriate packet headers to achieve internal or edge delivery.

In [32] and [33] the authors propose a generalization of SDN/OpenFlow architecture to include content based primitives related to storage and caching. Content routing is supported by mapping a content to a TCP flow and using an SDN approach to route the flow. The mapping is performed by edge proxies that map HTTP requests into flows. The main difference with our approach is that no ICN protocol is used, but clients and server just use legacy HTTP. In [34] the work presented in [32] and [33] is extended by introducing content based traffic engineering features.

As for the use of "tagging" in ICN networks, an architecture based on such an approach has been proposed in [35] for routing within a Data Center.

The approach of having a centralized routing engine per autonomous system has also been considered for regular IP routing. For example [36] first proposed the so-called Routing Control Platforms, while the BGP "Route reflector" solution [37] can be seen as a form of centralization of RIBs. Recently, solutions to centralize the RIB avoiding to run BGP at the edges of an autonomous system and exploiting an SDN backbone has been proposed in [38] and [39].

As for the inter-domain name-based routing, the use of a BGP-like approach was already proposed in [40]. Recently, it has been proposed to extend BGP to disseminate advertisement information about content objects in the form of URIs [41].

## 9. Conclusions

In this paper we have discussed an architecture for the support of ICN using SDN concepts, in particular considering extensions to the OpenFlow protocol. While focused on a specific flavor of ICN, called CONET, most considerations and results are of general validity, in particular we point out that we are able to support the existing CCNx applications and, more in general, all ICN proposals which follow the basic way of operation of CCN/NDN. We have first illustrated the design of a "long term" solution bearing no limitations with respect to the evolution of OpenFlow switches and OpenFlow protocol. Then we have presented the design and implementation of a "short-term" solution that has been deployed over the OFELIA testbed, using OpenFlow v1.0 equipment.

Some results we find worth mentioning are listed hereafter:

1) If SDN/OpenFlow is used, ICN information is needed in every packet and IP fragmentation should be avoided, the segmentation approach we introduced with CONET is also useful for this purpose.

2) We had previously defined a way to transport ICN information with IP Options in the IPv4 and IPv6 headers [18]. Considering an SDN/OpenFlow based solution, it is simpler to transport this information in transport level headers (assuming one has a CONET transport protocol). The OpenFlow flow matching approach natively considers headers of different layers in the processing of packets.

3) It is not practical to perform matching over ICN names of variable length. This suggests either to natively use fixed length names in ICN (but this does not fit well all naming schema) or to use a "tagging" approach. The idea is to add tags when entering an SDN domain and remove them at the egress.

4) We have started the analysis of a Northbound API offering ICN functionality. We believe this is an interesting topic that needs to be more deeply analyzed.

## 10. Acknowledgements

This work was partly supported by the EU in the context of the FP7 projects CONVERGENCE [6], OFELIA [14] and GreenICN.

## 11. References

[1] T. Koponen, M. Chawla, B.G. Chun, et al.: "A data-oriented (and beyond) network architecture", ACM SIGCOMM 2007
[2] V. Jacobson, D. K. Smetters, J. D. Thornton et al., "Networking named content", ACM CoNEXT 2009
[3] Named Data Networking (NDN) project, http://www.named-data.net/
[4] PURSUIT project, http://www.fp7-pursuit.eu
[5] SAIL project, http://www.sail-project.eu/
[6] CONVERGENCE project, http://www.ict-convergence.eu
[7] COMET project http://www.comet-project.org
[8] CCNx project web site: http://www.ccnx.org
[9] N. McKeown, T. Anderson, H. Balakrishnan, G. Parulkar, L. Peterson, J. Rexford, S. Shenker, and J. Turner, "OpenFlow: Enabling Innovation in Campus Networks". White paper. March 2008 (available at: http://www.openflow.org).
[10] http://www.openflow.org/
[11] https://www.opennetworking.org/
[12] A. Detti, N. Blefari-Melazzi, S. Salsano, and M. Pomposini. "CONET: A Content-Centric Inter-Networking Architecture", Proc. of ACM Sigcomm – ICN 2011. August 2011
[13] A. Detti, M. Pomposini, N. Blefari-Melazzi, S. Salsano, "Supporting the Web with an Information Centric Network that Routes by Name", Elsevier Computer Networks, vol. 56, Issue 17, p. 3705–3722
[14] OFELIA project: http://www.fp7-ofelia.eu

[15] Md. Faizul Bari et al, "A Survey of Naming and Routing in Information-Centric Networks", *IEEE Communications Magazine*, December 2012
[16] S. Salsano, A. Detti, M. Cancellieri, M. Pomposini, N. Blefari-Melazzi, "Transport-layer issues in Information Centric Networks", ACM SIGCOMM Workshop on Information-Centric Networking (ICN-2012), Helsinki, Finland, August 2012
[17] S. Salsano, M. Cancellieri, A.Detti, "ICTP - Information Centric Transport Protocol for CONET ICN", draft-salsano-ictp-02, Work in progress, June 20, 2013
[18] A. Detti, S. Salsano, N. Blefari-Melazzi, "IP protocol suite extensions to support CONET Information Centric Networking", Internet Draft, draft-detti-conet-ip-option-05, Work in progress, June 20, 2013
[19] B.Carpenter, S. Brim "Middleboxes: Taxonomy and Issues", IETF RFC 3234, February 2002
[20] L. Veltri, G. Morabito, S. Salsano, N. Blefari-Melazzi, A. Detti, "Supporting Information-Centric Functionality in Software Defined Networks", SDN'12: Workshop on Software Defined Networks, co-located with IEEE ICC, June 10-15 2012, Ottawa, Canada
[21] K Katsaros, G. Xylomenos, G. C. Polyzos: "MultiCache: An overlay architecture for information-centric networking", Computer Networks, Elsevier, Volume 55, Issue 4, 10 March 2011, Pages 936-947
[22] D. Smetters, V. Jacobson: "Securing Network Content", PARC technical report, October 2009
[23] "OpenFlow Switch Specification", Version 1.3.0 (Wire Protocol 0x04) June 25, 2012, Open Networking Foundation
[24] S. Salsano (editor),"EXOTIC final architecture and design", Deliverable D9.1, Project FP7 258365 "OFELIA", http://netgroup.uniroma2.it/Stefano_Salsano/papers/OFELIA_D9.1_v1.0.pdf
[25] Floodlight OpenFlow Controller Home Page, http://floodlight.openflowhub.org/
[26] A. Köpsel and H. Woesner, "OFELIA – Pan-European Test Facility for OpenFlow Experimentation", Lecture Notes in Computer Science. Vol. 6994/2011. 2011
[27] Enterprise GENI (eGENI) project: http://groups.geni.net/geni/wiki/EnterpriseGeni
[28] N. Blefari-Melazzi, A. Detti, G. Mazza, G. Morabito, S. Salsano, L. Veltri, "An OpenFlow-based Testbed for Information Centric Networking", Future Network & Mobile Summit 2012, 4-6 July 2012, Berlin, Germany.
[29] D. Syrivelis, G. Parisis, D. Trossen, P. Flegkas, V. Sourlas, T. Korakis, L. Tassiulas, "Pursuing a Software Defined Information-centric Network", EWSDN 2012, Darmstadt, Germany
[30] D. Trossen, G. Parisis. "Designing and Realizing an Information-Centric Internet" in IEEE Communications Magazine - Special Issue on Information-centric Networks, July 2012
[31] B. Raghavan, M. Casado, T. Koponen, S. Ratnasamy, A. Ghodsi, S. Shenker, "Software-defined internet architecture: decoupling architecture from infrastructure", 11th ACM Workshop on Hot Topics in Networks (HotNets-XI), October 29-30, 2012, Redmond, WA
[32] A. Chanda, C. Westphal, "Content as a network primitive", arXiv preprint arXiv:1212.3341 (2012).
[33] A. Chanda, C. Westphal, "ContentFlow: Mapping Content to Flows in Software Defined Networks", arXiv preprint arXiv:1302.1493 (2013).
[34] A. Chanda, C. Westphal, D. Raychaudhuri, "Content Based Traffic Engineering in Software Defined Information Centric Networks", 2nd IEEE International Workshop on Emerging Design Choices in Name-Oriented Networking (NOMEN 2013), Turin, Italy, April 2013.
[35] B. J. Ko, V. Pappas, R. Raghavendra, Y. Song, R. B. Dilmaghani, K.-won Lee, D. Verma, "An information-centric architecture for data center networks". 2nd ICN workshop on Information-centric networking (ICN '12), Helsinki, Finland
[36] M. Caesar, D. Caldwell, N. Feamster, J. Rexford, A. Shaikh, and J. van der Merwe, "Design and implementation of a routing control platform", in NSDI'05, 2005
[37] T. Bates, E. Chen, R. Chandra "BGP Route Reflection:An Alternative to Full Mesh Internal BGP (IBGP)", IETF RFC 4456, April 2006
[38] Christian Esteve Rothenberg, Marcelo Ribeiro Nascimento, Marcos Rogerio Salvador, Carlos Nilton Araujo Corrêa, Sidney Cunha de Lucena, and Robert Raszuk, "Revisiting Routing Control Platforms with the Eyes and Muscles of Software-Defined Networking", First workshop on Hot topics in software defined networks (HotSDN '12), 2012, Helsinki, Finland

[39] V. Kotronis, X. Dimitropoulos, B. Ager, "Outsourcing the routing control logic: better internet routing based on SDN principles", 11th ACM Workshop on Hot Topics in Networks (HotNets-XI), 2012, New York, USA

[40] M. Gritter, D. Cheriton, "An Architecture for Content Routing Support in the Internet", Proc. Usenix USITS, March 2001

[41] A. Narayanan, Ed., S. Previdi, B. Field, "BGP advertisements for content URIs", draft-narayanan-icnrg-bgp-uri-00, Work in progress, July 28, 2012